\documentclass[
  english,aps,prstper,reprint,showpacs,titlepage,
  longbibliography,superscriptaddress
]{revtex4-2}
\usepackage[T1]{fontenc}	
\usepackage{geometry}		
\usepackage{graphicx}
\usepackage[above,below]{placeins}	
\usepackage{times}
\usepackage{lineno}
\usepackage{booktabs}
\usepackage{siunitx}
\usepackage{multirow}

\usepackage{hyperref}  
\hypersetup{colorlinks=true,urlcolor=blue,citecolor=blue,linkcolor=blue}   
\usepackage{enumitem}          
\setlist{nosep}                 

\begin{document}
\begin{titlepage}

\title{Active Learning with Generative AI in Introductory Physics: A Multi-Institutional Study of the P-A.I.R. Framework}
\author{Bilas Paul}
\email{palb@farmingdale.edu}
\affiliation{Department of Physics, SUNY Farmingdale State College, Farmingdale, NY 11735, USA}
\author{Shantanu Chakraborty}
\affiliation{Department of Physics, Valdosta State University, Valdosta, GA 31698, USA}
\author{Shruti Shrestha}
\affiliation{Department of Physics, Penn State Brandywine, Media, PA 19063, USA}
\author{Binod Nainabasti}
\affiliation{Department of Physics, Lamar University, Beaumont, TX 77705, USA}
\author{Bisaj Paul}
\affiliation{Connetquot High School, Bohemia, NY 11715, USA}

\begin{abstract}
\label{abstract}
\noindent
Generative AI tools have become part of how students study physics, yet unstructured use may reduce students’ opportunities to work through problems independently. The Physics AI-Replication (P-A.I.R.) framework structures AI use as a four-step cycle: students identify a challenging problem, obtain an AI-generated explanation and solution, request similar practice problems, and solve them independently before reviewing the AI's solutions. We evaluated P-A.I.R.\ in 11 algebra- and calculus-based introductory physics sections at four U.S.\ institutions using a common end-of-semester survey. Most students reported improved conceptual understanding (90.9\%) and found the replicated practice problems helpful (97.7\%), and these two ratings were strongly associated ($r_s = 0.58$). Perceptions of both components, as well as post-engagement confidence ($\mu = 7.18$ on a 1--10 scale), were consistent across institutions. Post-engagement confidence was associated with perceived improvement in conceptual understanding ($r_s = 0.37$) and the helpfulness of replicated practice ($r_s = 0.35$), but not with how often students used AI ($r_s = 0.02$) or with self-reported GPA ($r_s = 0.13$). Thematic analysis of open-ended responses showed that students valued AI primarily for conceptual clarification and structured problem solving (63.0\%), while their main concerns were AI accuracy (34.2\%) and the need for explanations that guide rather than simply provide answers (29.7\%). These findings suggest that how students engage with AI matters more than how often they use it, and that frameworks sequencing AI assistance with independent problem solving offer a practical, transferable approach for physics instruction.
\end{abstract}

\maketitle

\end{titlepage}

\section{Introduction }
\label{intro}
Generative artificial intelligence (AI) systems are rapidly becoming part of students’ everyday academic practices. Large language models such as ChatGPT and Gemini can explain concepts, generate worked solutions, create practice problems, and provide immediate feedback, offering students a form of continuously available academic support~\cite{Kasneci2023ChatGPTFG, baidoo2023education}. The speed of this adoption has outpaced the development of pedagogical norms governing it: students are already integrating these tools into their study routines, often without guidance on how to do so productively. As these tools become increasingly accessible, physics instructors face an important educational question: how can AI use support meaningful learning and independent reasoning?

The growing capabilities of generative AI make the question of how students engage with these tools increasingly consequential for physics instruction. Although large language models (LLMs) can perform successfully on many physics tasks, their responses are not uniformly reliable.  Our recent evaluation of generative AI systems on algebra-based AP Physics free-response questions showed that their performance varies across problem types and that AI-generated solutions can contain errors in physical reasoning and problem solving \cite{Paul2026APAI}. Other studies have similarly identified both opportunities and limitations in the use of generative AI for physics learning \cite{Liang2023ExploringTP,Kieser2023}. 

These concerns are particularly relevant in introductory physics, where successful learning depends not only on obtaining correct numerical answers but also on developing conceptual reasoning, problem-solving strategies, and the ability to transfer knowledge across contexts. Expertise in physics is characterized less by the recall of isolated facts than by the organization of knowledge into coherent conceptual structures that support flexible application~\cite{chi1981categorization}. Prior research in physics education has consistently shown that active engagement, deliberate practice, and constructive interaction are central to durable learning~\cite{hake1998interactive, freeman2014active, chi2009active}.  The cognitive science literature offers complementary
mechanisms that explain why such engagement matters. Retrieval practice---the act of reconstructing knowledge from memory rather than rereading it---produces more durable learning than repeated exposure~\cite{roediger2006test}, and the generation effect indicates that information a learner produces is better retained than information passively received~\cite{slamecka1978generation}. Unrestricted AI use may unintentionally encourage passive dependence on generated solutions, reducing opportunities for productive struggle and reflective reasoning~\cite{kirschner2006minimal}. When a complete, polished solution is always one prompt away, the learner is rarely required to retrieve, generate, or struggle---the very activities that consolidate understanding. Consequently, there is growing interest in identifying instructional structures that guide students toward more educationally productive uses of AI.  

The educational implications of generative AI therefore depend not simply on whether students use these tools, but on the cognitive activities that AI use replaces or elicits. During physics problem solving, students ordinarily must interpret the problem, identify relevant physical principles, construct or select representations, carry out a solution, and check the consistency and correctness of the solution. Generative AI can support several of these processes, but it can also reduce the need for students to carry out these steps independently. The same technological capability may therefore function either as a scaffold for learning or as a substitute for cognitive engagement, depending on how the student--AI interaction is structured.

A useful lens for distinguishing such uses is the Interactive--Constructive--Active--Passive (ICAP) framework, which orders learning activities from passive, through active and constructive, to interactive, predicting that more generative forms of engagement yield deeper learning~\cite{chi2014icap}.  Through this lens, simply reading an AI-generated solution is a passive activity, whereas attempting a novel problem before consulting the solution is constructive, and iteratively comparing one's own reasoning against the AI's feedback approaches the interactive end of the spectrum. The pedagogical challenge, then, is to design interactions that move students up this hierarchy rather than anchoring them at its passive base. From the perspective of cognitive load theory, worked examples can be valuable precisely because they free working-memory resources for schema construction~\cite{sweller1988cognitive};  the difficulty is ensuring that students study those examples generatively rather than merely glancing at the final result.

A complementary perspective is provided by self-regulated learning, in which learners actively plan, monitor, and evaluate their own learning processes. AI systems have the potential to support such regulation by providing immediate explanations, feedback, and opportunities for additional practice, but they may also shift parts of the regulatory process from the learner to the technology. A recent systematic review of AI-supported self-regulated learning in higher education highlights the potential of AI tools to support learner autonomy while also emphasizing the importance of pedagogical designs that preserve active learner regulation~\cite{LanZhou2025}. In the context of physics, this distinction suggests that asking an AI system for a completed solution and reading it is fundamentally different from using the system to generate a related problem, attempting that problem independently, and subsequently using the AI response for comparison and reflection.

Emerging physics education research further suggests that students themselves recognize this tension. Recent work on physics students' adoption of ChatGPT revealed contrasting patterns of use. Some students used generative AI
pragmatically for purposes such as conceptual clarification and obtaining starting points for problem solving, whereas others avoided the technology because of concerns about over-reliance and the possible erosion of independent problem-solving skills~\cite{Becker2026}. Such findings shift the instructional question away from a simple choice between allowing and prohibiting AI. Instead, they motivate the development and evaluation of pedagogical structures that specify how AI should participate in the learning process while preserving substantial cognitive responsibility for the student.

To address this need, we previously introduced the Physics AI-Replication (P-A.I.R.) framework, a structured approach designed to encourage active interaction with AI-generated explanations and practice problems~\cite{Paul2026PAIR}. Within this framework, students first identify a difficult physics problem, request conceptual explanations and worked solutions from an AI system, generate analogous practice problems, and then attempt those problems independently before reviewing the AI-generated solutions. This sequence is theoretically motivated: the explanation phase supplies a worked example that manages cognitive load, while the independent-attempt phase reintroduces retrieval and generation demands before the student is permitted to check the answer. Rather than treating AI as a direct answer provider, the framework positions it as a scaffold for iterative practice, conceptual clarification, and self-guided learning, encouraging students to move  from the passive to the constructive end of the engagement spectrum.

A preliminary single-institution study of the P-A.I.R. framework suggested that students perceived the framework positively, particularly in relation to conceptual understanding, confidence, and problem-solving support~\cite{Paul2026PAIR}. However, the generalizability of those findings remained limited by the relatively small sample size and single-campus context. Student populations, instructional environments, and prior familiarity with AI technologies may vary substantially across institutions, potentially influencing how structured AI-supported learning is experienced.

More broadly, although the literature on generative AI in physics education is expanding rapidly, much of the existing work has focused on the capabilities and limitations of AI systems themselves \cite{Liang2023ExploringTP, Kieser2023, Paul2026APAI}, on relatively small or single-context implementations, or on students' perceptions and patterns of AI adoption~\cite{Bitzenbauer2023, Becker2026}. Less is known about whether a theoretically motivated
instructional framework that sequences AI assistance with independent problem solving produces similar student experiences across institutions, course types, and local implementations. Multi-institutional evidence is particularly important for distinguishing patterns associated with the instructional framework from those that may arise from a particular instructor, course environment, or student population.

The present study expands this preliminary investigation by incorporating data from three additional institutions into the original dataset. Using a common survey instrument and shared implementation framework, we examine students’ perceptions of conceptual learning, usability, confidence, efficiency, and continued adoption of the P-A.I.R. process across diverse instructional settings. 

Accordingly, the present study addresses the following research questions:

\begin{enumerate}
    \item[RQ1:] How do introductory physics students perceive the conceptual, practice-related, and usability benefits of the P-A.I.R. framework?

    \item[RQ2:] To what extent are students' perceptions of the core components of P-A.I.R. consistent across institutions and instructional contexts?

    \item[RQ3:] How are students' prior AI use and confidence in using   AI, their experiences with the core components of P-A.I.R.,  and their subsequent confidence in solving physics problems independently related?
\end{enumerate}

By examining these questions across multiple institutions and course environments, this study investigates whether the perceived value of structured AI-supported learning extends beyond a single classroom context and, more broadly, whether the manner in which students engage with generative AI may be more educationally consequential than the frequency with which they use it.

\section{Materials and Methods}
\label{method}
\begin{table*}[htb!]
  \centering
  \caption{Participating institutions, course sections, enrollment, and
    survey participation. Course types are abbreviated as
    alg.\ (algebra-based) and calc.\ (calculus-based); Physics~I denotes
    introductory mechanics and Physics~II denotes electricity and
    magnetism. Participation is the percentage of enrolled students who
    completed the survey.}
  \label{tab:sites}
  \begin{ruledtabular}
    \begin{tabular}{l l c  c c}
      Institution & Course sections 
        & Enrolled & Surveyed & Participation (\%) \\
      \colrule
      SUNY Farmingdale State College (FSC)\footnotemark[1]
        & 2$\times$ Physics~I (alg.), 1$\times$ Physics~II (alg.) 
        & 72  & 39 & 54.2 \\
      Valdosta State University (VSU)
        & 2$\times$ Physics~I (alg.), 2$\times$ Physics~II (calc.) 
        & 122 & 71 & 58.2\\
      Penn State Brandywine (PSU-BW)
        & 2$\times$ Physics~II (calc.), 1$\times$ Physics~I (alg.) 
        & 100 & 73 & 73.0 \\
      Lamar University (LU)
        & 1$\times$ Physics~I (alg.) 
        & 64  & 52 & 81.3 \\
      \colrule
      Total & 11 sections & 358 & 235 & 65.6 \\
    \end{tabular}
  \end{ruledtabular}
  \footnotetext[1]{Data previously reported in the single-institution conference study~\cite{Paul2026PAIR}.}
\end{table*}

This study employed a multi-institutional survey-based research design to evaluate students' perceptions of the P-A.I.R. framework across diverse instructional settings. Building on the preliminary single-campus study~\cite{Paul2026PAIR}, the present work expands the original dataset by incorporating data from three additional institutions. This extension allowed us to examine whether perceptions of structured AI-supported learning persist beyond a single classroom environment. The study was conducted at four institutions in the United States during the 2025--2026 academic year: SUNY Farmingdale State College (FSC), Farmingdale, NY; Valdosta State University (VSU), Valdosta, GA;  Penn State Brandywine (PSU-BW), Media, PA; and Lamar University (LU), Beaumont, TX. To ensure consistency across institutions, participating instructors implemented a common P-A.I.R. protocol and administered an identical end-of-semester survey instrument, while adapting implementation details to local course structures and assessment practices.

Across the four campuses, participants were enrolled in introductory algebra-based and calculus-based physics courses spanning both first-semester (mechanics) and second-semester (electricity and magnetism) content. Table~\ref{tab:sites} summarizes the course sections, enrollment, and survey participation at each institution. In aggregate, \num{358} students were enrolled across \num{11} course sections, and \num{235} students completed the survey, corresponding to an overall response rate of 65.6\%.

All study procedures were reviewed and approved by the Institutional Review Board (IRB)  at each participating institution. At every site, informed consent was obtained from all participants, who were assured of the anonymity and confidentiality of their responses and of their right to withdraw at any time without penalty. Survey data were stored securely and were not accessed by the principal investigator at any institution until after final course grades had been submitted, ensuring that participation could not influence students' academic evaluation.

The P-A.I.R. framework is a structured approach designed to encourage active engagement with generative AI tools during physics problem solving. The framework was introduced near the beginning of the semester and remained available throughout the course. Students were encouraged, but not required, to apply the protocol whenever they encountered challenging homework, practice, recitation, or examination-preparation problems. Each use of the framework involved four sequential steps: (1)~identifying a challenging problem from course materials; (2)~using an AI tool of their choice (e.g., ChatGPT, Gemini, Copilot, or similar systems) to obtain a conceptual explanation and step-by-step solution; (3)~requesting one or two new, similar practice problems from the AI; and (4)~attempting those problems independently before reviewing the AI-generated solutions. The framework was designed to move students beyond simply viewing AI-generated solutions by incorporating problem replication and independent practice into the learning process. Although implementation details varied somewhat across institutions to accommodate local course structures and assessment practices, the four-step P-A.I.R. protocol remained unchanged.

At the end of the semester, after students had sufficient opportunity to engage with the P-A.I.R. framework, participants were invited to complete a voluntary survey. At FSC, the survey was delivered via Qualtrics with IP-address collection disabled. At VSU, PSU-BW, and LU, anonymous paper-based surveys were administered in class and collected by the end of the class period and submitted to the principal investigator.  Survey content and response options were identical across institutions. The instrument was designed to capture students' educational backgrounds, their experiences with the P-A.I.R. framework, and their perceived learning outcomes, and comprised three sections. \emph{Section~A} collected background information, including the frequency and primary purposes of AI use in physics coursework, self-assessed confidence in using AI effectively, anticipated GPA range, intended major, and the specific AI tools used. \emph{Section~B} contained seven experience items assessed on ordinal Likert-type scales: (B1)~self-assessed improvement in conceptual understanding (4-point); (B2)~helpfulness of the AI-generated replicated practice problems (4-point); (B3)~ease of interaction with AI tools (5-point); (B4)~likelihood of continued AI use (5-point); (B5)~perceived reliability and accuracy of AI-generated responses (4-point); (B6)~time efficiency relative to traditional study methods (5-point); and (B7)~likelihood of recommending the P-A.I.R. approach to peers (5-point). \emph{Section~C} comprised one quantitative and two qualitative items: (C1)~a self-rated assessment of problem-solving confidence on a 1--10
scale; (C2)~an open-ended prompt asking students to describe one specific way in which the P-A.I.R. framework influenced their approach to studying physics; and (C3)~an open-ended prompt soliciting suggestions for improving
the P-A.I.R. process.

Quantitative survey responses were analyzed using descriptive statistical techniques implemented in Python to examine trends in student perceptions across items. Given the ordinal nature of the Likert-scale data, non-parametric statistical tests were employed where appropriate. Qualitative responses to the open-ended questions were analyzed using an inductive thematic analysis approach~\cite{braun2006using}. Each participating instructor first coded the responses from their own course sections, and an independent researcher then coded all responses from the four institutions using a common codebook. Agreement between the two codings was assessed for each student whose responses were included in the qualitative analysis. Disagreements in code assignments were resolved through discussion, and the reconciled codes were used in the analysis. Representative excerpts are reported with minor typographical corrections where necessary to improve readability.

\section{Results}
\label{results}
Of the 358 students enrolled across the four participating institutions, 235 completed the end-of-semester survey, corresponding to an overall response rate of 65.6\%, although not all respondents answered every survey item. Participants represented a broad range of academic disciplines. Two similarly sized groups comprised the majority of the sample: students in pre-medical, biological, and health science disciplines (43.8\%, n = 103) and students from engineering- and technology-related disciplines—including mechanical, civil, electrical, architectural, and chemical engineering, engineering technology, computer science, and cybersecurity (43.0\%, n = 101). The remaining participants represented chemistry, liberal arts, and other disciplines.

AI tools were already widely used for academic purposes prior to the implementation of the P-A.I.R. framework. More than one-third of students (34.9\%) reported using AI often (2–3 times per week), 36.2\% reported occasional use (approximately once per week), and 7.7\% reported daily use, whereas only 17.0\% reported rare use and 3.0\% indicated that they never used AI for academic work. ChatGPT was the most commonly used platform (68.5\%), followed by Google Gemini (43.8\%) and Microsoft Copilot (22.1\%), with many students reporting the use of multiple AI systems. Students most frequently used AI for concept explanation (82.6\%), followed by problem-solving assistance (54.9\%), exam preparation (48.5\%), and homework support (39.1\%). Despite this widespread adoption, students' self-reported confidence in using AI effectively for physics remained moderate overall: 44.3\% reported moderate confidence, 27.7\% reported high confidence, and only 4.3\% reported extreme confidence, while 17.0\% reported slight confidence and 5.5\% reported no confidence at all. AI usage frequency was positively correlated with students' self-rated confidence in using AI effectively in introductory physics courses (Spearman's $r_s = 0.47, p < 0.001, n = 231$), indicating that students who used AI tools more frequently also tended to report greater confidence in using them effectively.

\begin{figure}[t]
\centering
\includegraphics[width=0.95\columnwidth]{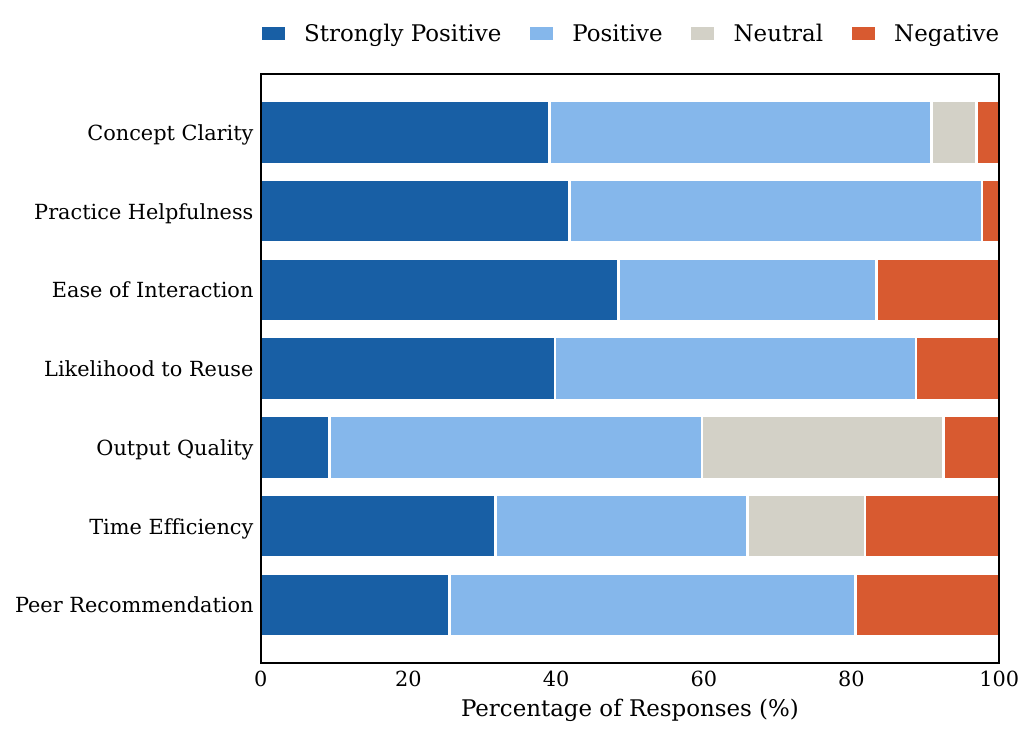}
\caption{Distribution of student responses to the seven \textit{Section~B} experience items (B1--B7), pooled across the four participating institutions. Responses are grouped into four sentiment bands and are strongly skewed toward positive categories, with particularly high ratings for conceptual understanding (B1) and problem-replication helpfulness (B2). Output quality (B5) shows the largest neutral share, reflecting a pragmatic recognition of the limitations of generative
AI.}
\label{fig:likert}
\end{figure}

\begin{figure}[t]
\centering
\includegraphics[width=0.95\columnwidth]{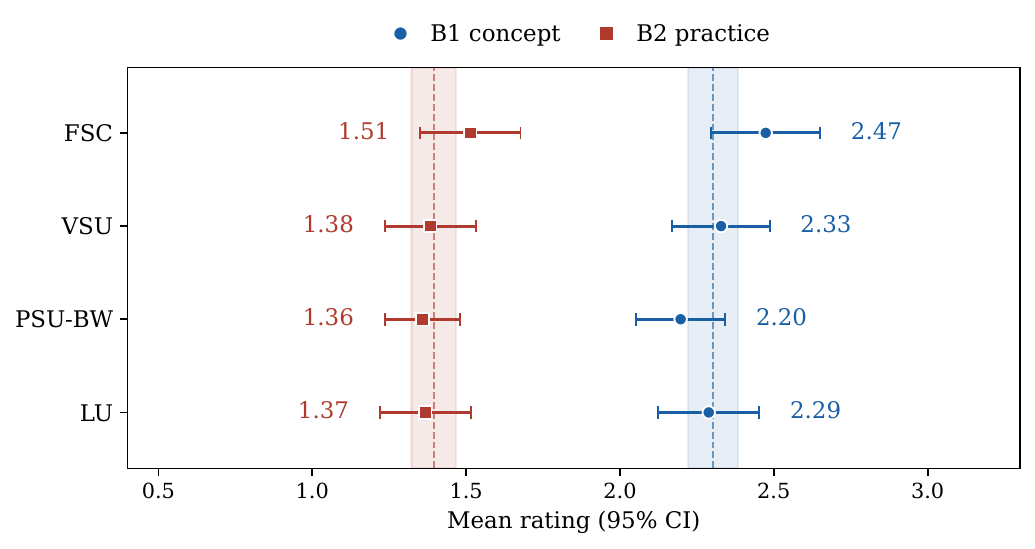}
\caption{Mean ratings by institution for the two core P-A.I.R. experience items: conceptual understanding (B1, blue circles) and replicated practice (B2, red squares). Error bars represent $95\%$ confidence intervals for the institutional means. Dashed lines indicate the respective pooled means, and shaded bands show their $95\%$ confidence intervals. Institutional confidence intervals overlap the pooled mean for both items, suggesting broadly similar perceptions across all four institutions.}
\label{fig:forestb1b2}
\end{figure}

\begin{figure}[t]
\centering
\includegraphics[width=\columnwidth]{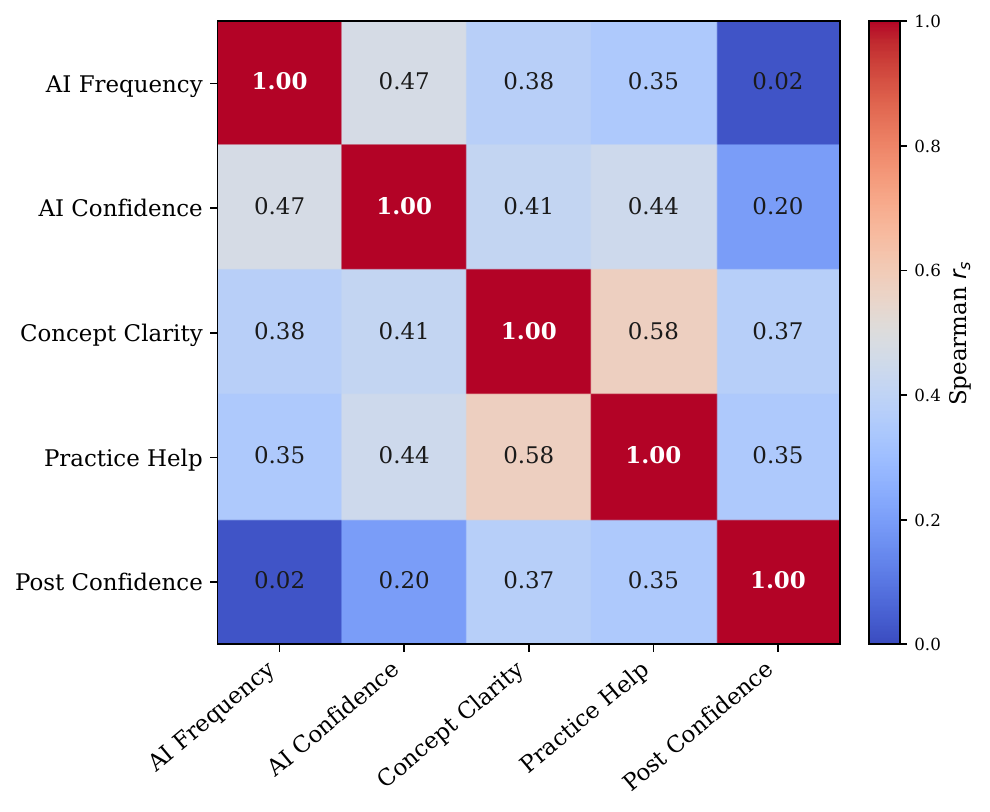}
\caption{Spearman rank correlation matrix among AI usage frequency, prior AI confidence, conceptual understanding (B1), helpfulness of replicated practice (B2), and post-engagement self-confidence (C1). All pairwise associations among the first four variables are positive and significant ($p < 0.001$); the strongest is between B1 and B2 ($r_s = 0.58$). Post-engagement confidence (bottom row) is most strongly associated with the two experience items (B1, B2) and shows a negligible, non-significant association with AI usage frequency.}
\label{fig:heatmap}
\end{figure}

Student responses to the seven \textit{Section~B} experience items were strongly positive across all four institutions (Table~\ref{tab:sectionB}; Fig.~\ref{fig:likert}). 
\begin{table}[htb!]
\caption{Summary statistics for \textit{Section~B} experience items, pooled across the four participating institutions. \%~pos.\ denotes the percentage of valid responses at or above the positive threshold for each item. Items were measured on varying Likert-type scales and coded numerically for descriptive analysis.}
\label{tab:sectionB}
\begin{ruledtabular}
\begin{tabular}{llcccc}
 & Item & $n$ & $\mu$ & $\sigma$ & \%~pos. \\
\colrule
B1 & Concept clarity        & 231 & 2.30 & 0.63 & 90.9 \\
B2 & Practice helpfulness   & 213 & 1.39 & 0.54 & 97.7 \\
B3 & Ease of interaction    & 228 & 3.75 & 0.76 & 83.8 \\
B4 & Likelihood to reuse    & 231 & 3.81 & 0.80 & 88.7 \\
B5 & AI reliability and accuracy         & 226 & 2.62 & 0.76 & 59.7 \\
B6 & Time efficiency        & 220 & 3.75 & 1.16 & 65.9 \\
B7 & Peer recommendation    & 231 & 3.64 & 0.95 & 80.5 \\
\end{tabular}
\end{ruledtabular}
\end{table}
The two items most directly tied to the framework's core mechanism---improvement in conceptual understanding (B1) and the helpfulness of the AI-generated replicated practice problems (B2)---were the most uniformly endorsed. For B1, $90.9\%$ of respondents reported that their conceptual understanding had improved, with $39.4\%$ describing the improvement as significant and $51.5\%$ as moderate; only a small minority (6.1\%) reported no change, and $3.0\%$ indicated that AI use had made the underlying concepts more confusing. The replicated-practice component (B2) showed a similarly strong distribution, with $97.7\%$ of students rating it helpful ($41.8\%$ very helpful, $55.9\%$ somewhat helpful). Mean ratings for conceptual understanding and replicated practice were broadly similar across the four institutions, with overlapping $95\%$ confidence intervals (Fig.~\ref{fig:forestb1b2}).
The similarity of these response distributions is consistent with the sequential design of the P-A.I.R. framework, in which AI-generated explanations are intended to scaffold, rather than replace, students' subsequent independent practice. A Spearman rank-order correlation revealed a strong positive association between these two variables ($r_s = 0.58$, $p < 0.001$, $n = 213$). This relationship closely mirrors that observed in the original single-institution study~\cite{Paul2026PAIR}, suggesting that the link between conceptual understanding and productive practice is robust across diverse institutional settings rather than being unique to a particular classroom. To examine these two experience items in the context of students' broader engagement with AI, we analyzed their relationships with the two background measures from \textit{Section~A}: AI usage frequency and self-rated confidence in using AI. The resulting Spearman correlation matrix (Fig.~\ref{fig:heatmap}), which also includes post-engagement confidence (C1; discussed later),  shows that all four variables were positively and significantly interrelated.  Both perceived conceptual improvement and the helpfulness of replicated practice were more strongly associated with students' self-rated AI confidence ($r_s = 0.41$ and $0.44$, respectively; both $p < 0.001$) than with AI usage frequency ($r_s = 0.38$ and $0.35$, respectively; both $p < 0.001$), suggesting that students' perceived learning benefits were more closely related to their confidence in using AI effectively than to how frequently they used AI.

Responses to the remaining experience items (B3--B7) also indicated generally positive perceptions of the P-A.I.R. framework, although responses were more varied than those for conceptual understanding and replicated practice (Table~\ref{tab:sectionB}; Fig.~\ref{fig:likert}). Most students found the AI tools easy to interact with (B3), with $83.8\%$ rating them as somewhat, very, or extremely easy, including $48.7\%$ who selected very or extremely easy. Ratings of AI reliability and accuracy (B5) were more mixed: $59.7\%$ of respondents rated AI-generated responses as good or excellent, whereas $33.2\%$ selected fair, highlighting that many students recognized the educational value of the framework while remaining aware of the limitations of generative AI.  Compared with traditional study methods, $65.9\%$ of respondents perceived the P-A.I.R. approach to be more time-efficient (B6), with $31.8\%$ indicating that it was much more efficient. Indicators of future adoption were also highly favorable. Overall, $88.7\%$ of students reported that they were likely to continue using the P-A.I.R. framework (B4), and $80.5\%$ indicated that they would recommend it to their peers (B7).  Moreover, students who rated the two core experience items more highly were also more likely to intend future use of the framework: both conceptual clarity (B1) and replication helpfulness (B2) were significantly  associated with likelihood of continued use (B4; $r_s = 0.39$ and $0.29$, respectively) and with willingness to recommend the approach to peers (B7; $r_s = 0.42$ and $0.31$, respectively; all $p < 0.001$). This indicates that students who perceived the greatest learning benefit from P-A.I.R. were also those most inclined to continue using it and to endorse it to others.

\begin{figure}[t]
\centering
\includegraphics[width=0.95\columnwidth]{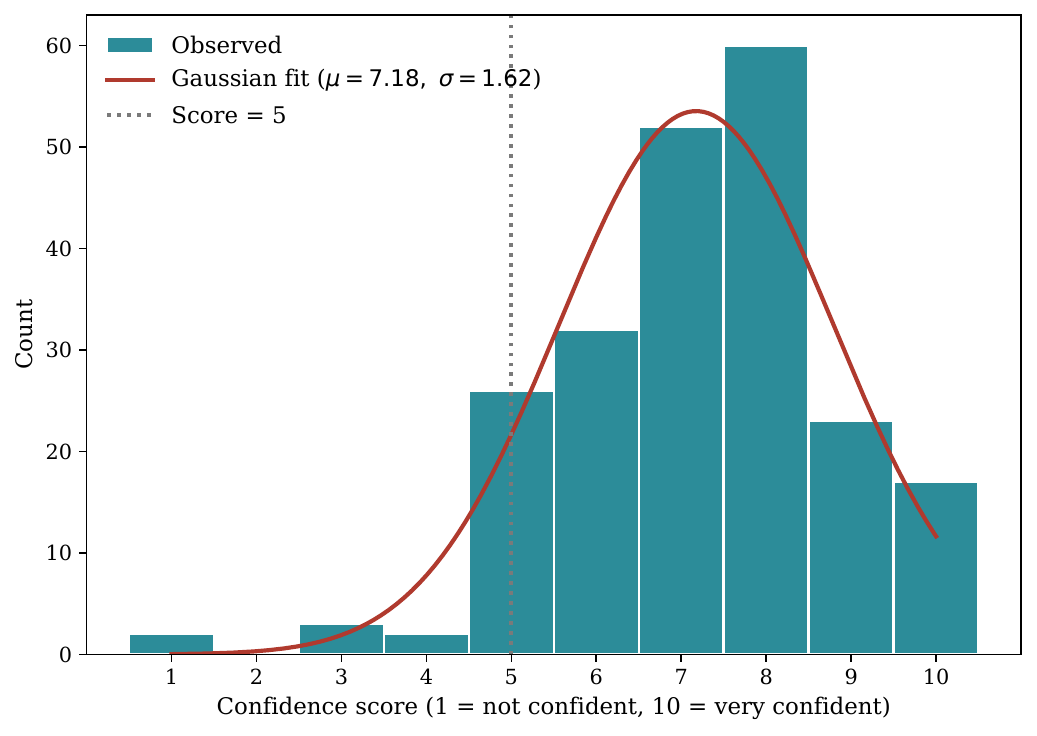}
\caption{Distribution of post-engagement self-confidence scores (C1; 1--10 scale), pooled across the four institutions. Scores are concentrated above 5 (dotted line), with only $3.7\%$ of responses below 5, indicating generally high perceived ability to solve similar problems independently after engaging with the P-A.I.R. framework. }
\label{fig:c1dist}
\end{figure}

\begin{figure}[t]
\centering
\includegraphics[width=0.95\columnwidth]{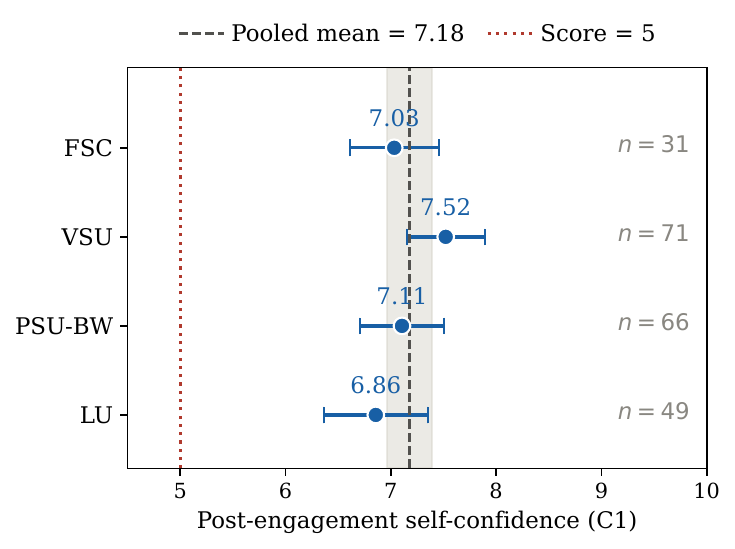}
\caption{Mean post-engagement self-confidence (C1) by institution, with $95\%$ confidence intervals. Institutional means (points) range from 6.86 to 7.52 and all lie well above 5 (dotted line); the shaded band and dashed line denote the pooled mean ($7.18$) and its $95\%$ confidence interval. The overlap of all four intervals indicates a consistent pattern across institutions.}
\label{fig:c1inst}
\end{figure}

Post-engagement self-confidence (C1; 1--10 scale) was reported by 217 participants ($\mu = 7.18$, $\sigma = 1.62$, $\mathrm{Mdn} = 7.0$; range 1--10). As shown in Fig.~\ref{fig:c1dist}, confidence scores were concentrated well above 5 on the 1--10 scale, with only 8 of 217 respondents ($3.7\%$) reporting a score below 5, indicating generally high self-reported confidence in solving similar problems independently after engaging with the P-A.I.R.\ framework. This level of confidence was consistent across all four participating institutions (Fig.~\ref{fig:c1inst}): institutional means ranged narrowly from 6.86 to 7.52, all well above the scale midpoint, with overlapping confidence intervals clustered around the pooled mean. The stability of this pattern across campuses with differing student populations and instructional settings indicates that the confidence associated with P-A.I.R. is not specific to a single institutional context.

To identify which aspects of the P-A.I.R.\ experience were associated with this confidence, we computed Spearman rank correlations between C1 and four candidate predictors (Fig.~\ref{fig:heatmap}). The two framework-experience items were the strongest correlates: perceived conceptual clarity (B1; $r_s = 0.37$, $p < 0.001$, $n = 216$) and the helpfulness of replicated practice (B2; $r_s = 0.35$, $p < 0.001$, $n = 198$). Prior confidence in using AI showed a weaker but still significant association ($r_s = 0.20$, $p = 0.004$, $n = 216$). Notably, AI usage frequency was essentially uncorrelated with post-engagement confidence ($r_s = 0.02$, $p = 0.749$, $n = 216$): how often students used AI tools was unrelated to how confident they felt solving problems independently afterward. This pattern suggests that students' confidence was more closely associated with their perceived benefits from explanation and replicated practice than with their reported frequency of AI use, consistent with the premise that the structure of AI-assisted study may matter more than frequency alone.

\begin{figure}[t]
\centering
\includegraphics[width=0.95\columnwidth]{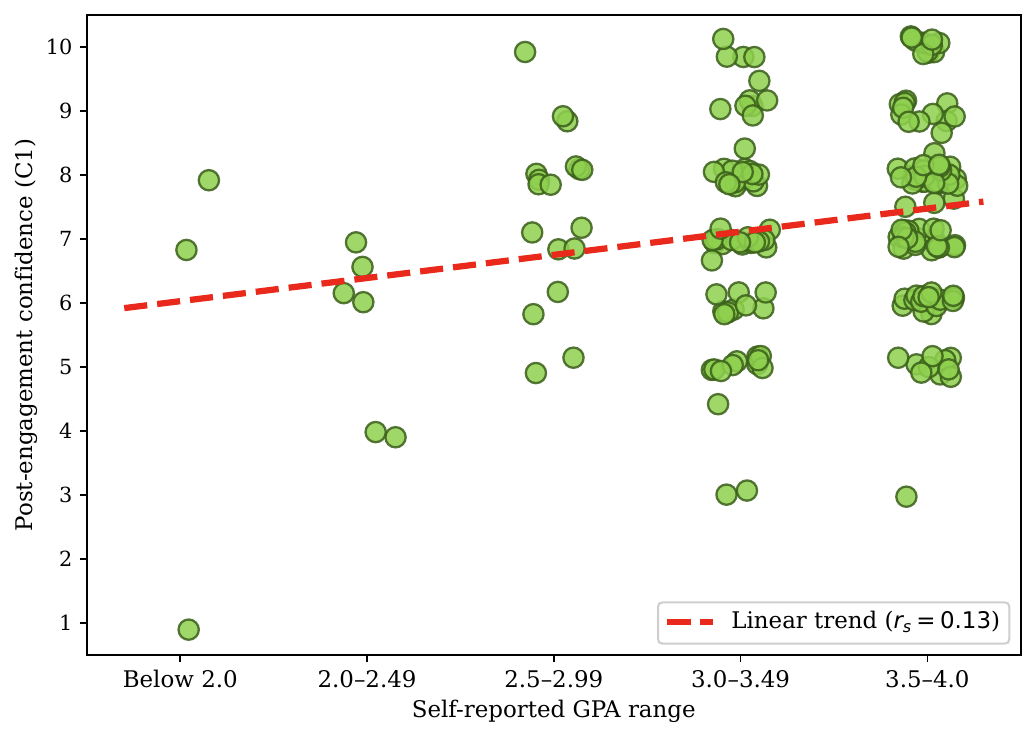}
\caption{Post-engagement self-confidence (C1) versus self-reported GPA range. Each point represents an individual respondent; the dashed line shows a fitted linear trend; the Spearman rank correlation is $r_s = 0.13$. Neither this association ($p = 0.064$) nor a Kruskal--Wallis test across three GPA groups (below 3.0, 3.0--3.49, and 3.5--4.0) ($H = 3.06$, $p = 0.217$) is statistically significant, indicating that post-engagement confidence did not differ reliably with academic standing.}
\label{fig:gpa}
\end{figure}

We examined the association between post-engagement confidence and students’ self-reported GPA range as a broad indicator of academic standing. Post-engagement confidence was not significantly associated with GPA ($r_s = 0.13$, $p = 0.064$, $n = 202$, Fig.~\ref{fig:gpa}), and a Kruskal--Wallis test across three GPA groups (below 3.0, 3.0--3.49, and 3.5--4.0) likewise revealed no significant differences in confidence ($H = 3.06$, $p = 0.217$). Students across the GPA groups reported confidence in solving similar problems independently after using P-A.I.R., suggesting that favorable confidence ratings were not limited to high-achieving students.

\subsection{Students' accounts of how P-A.I.R.\ shaped their study practices}
\label{sec:qual}
 
\begin{figure}[htb!]
    \centering
    \includegraphics[width=\columnwidth]{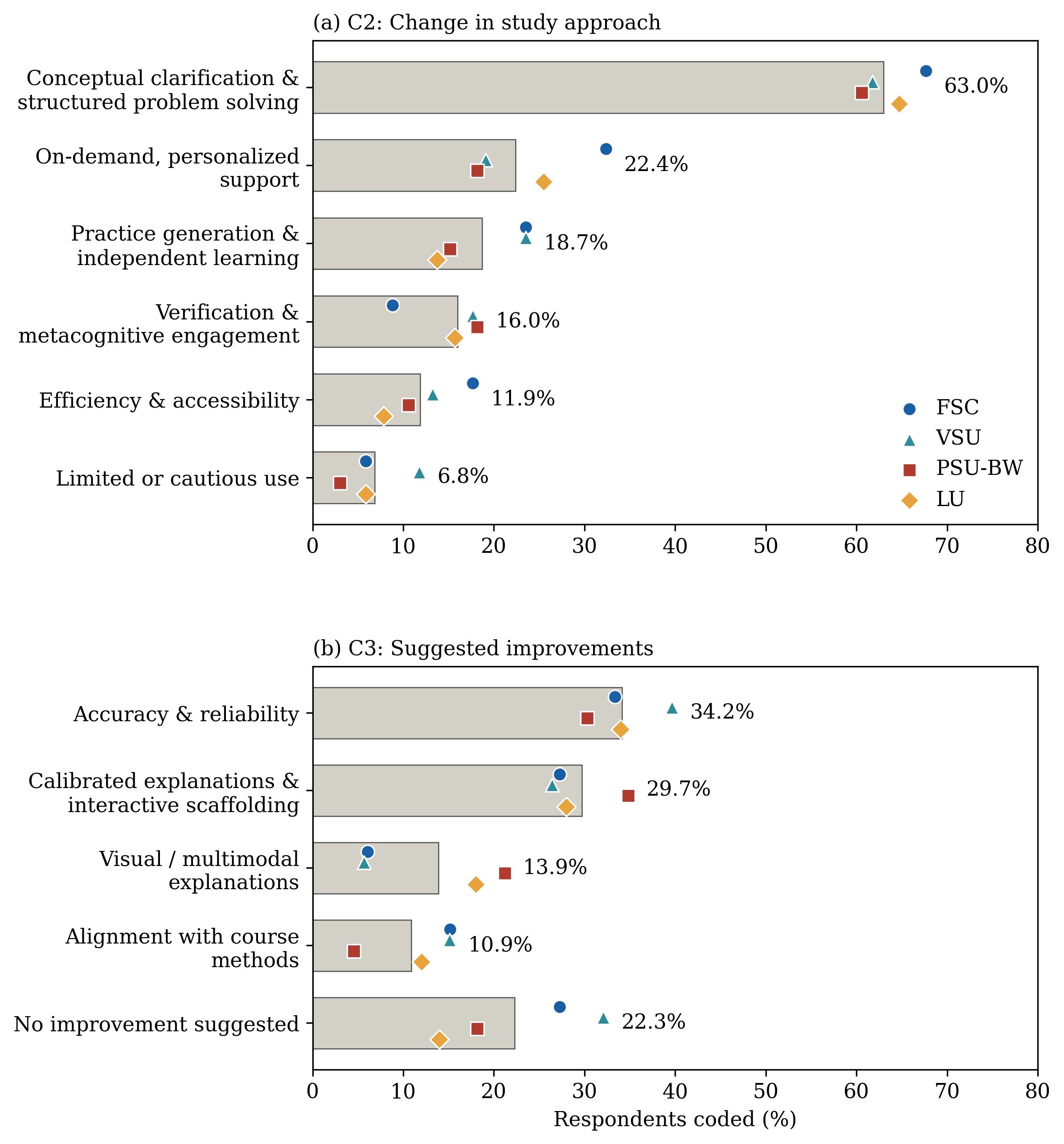}
    \caption{Prevalence of themes in students' open-ended responses: (a) how the P-A.I.R.\ process changed their approach to studying physics (C2) and (b) suggested improvements (C3). Bars show the pooled percentage of respondents whose response was coded with each theme; markers show the corresponding percentage at each institution.}
    \label{fig:themes}
\end{figure}

The quantitative results above show what students reported; the open-ended items (C2 and C3) offer insight into why. Of the 235 survey respondents whose responses were coded, 219 described how the P-A.I.R.\ process changed their approach to studying (C2) and 202 offered a suggestion for improving it (C3). Each response was independently coded by two researchers using six categories for C2 and five for C3. Multiple categories could be assigned to a response. Blank responses were coded as 0 (theme absent) for each applicable category by both coders.  Across 2,585 binary coding decisions for 235 students, initial percentage agreement was 86.8\%, with pooled Cohen's $\kappa=0.57$. Agreement was lowest for codes whose boundaries with neighboring codes required judgment about the student's intent, and highest for codes signaled by explicit wording in the response.  All disagreements were resolved through discussion before the frequencies reported below were computed. Figure~\ref{fig:themes} summarizes the prevalence of each theme, both pooled and by institution.

\textit{Changes in study approach (C2).}---Consistent with the high ratings for conceptual understanding (B1), the dominant theme was \textit{conceptual clarification and structured problem solving}, coded in 63.0\% of C2 responses. Its prevalence was remarkably stable across institutions, ranging from 60.6\% to 67.6\%. Students described AI as a means of breaking problems into manageable steps, deciding which equations applied, and understanding the reasoning behind a solution: \textit{``The AI tool explains step by step the process of solving questions and shows why I am supposed to do what I am doing''} (PSU-BW). This theme frequently co-occurred with \textit{on-demand, personalized learning support} (22.4\%), in which students treated AI as a patient, individual tutor. One student described \textit{``a personal tutor that I can ask deep contextual questions repeatedly without feeling annoying to someone''} (LU). Another described a deliberate questioning strategy: \textit{``I start by constantly asking `why' and `how' until I understand the root of the problem instead of just looking for an answer''} (VSU).

Two further themes map directly onto the replication and reflection stages of the framework. Students who reported \textit{practice generation and independent learning} (18.7\%) valued the ability to produce new problems on demand, often in the sequence the protocol prescribes: \textit{``If I get stuck on a HW problem, I use AI to show me the first step \ldots\ Then I ask it to give me more problems related to it so I can do it again''} (PSU-BW). Students coded for \textit{verification and metacognitive engagement} (16.0\%) described attempting problems before consulting AI, using it to locate their own mistakes, or critically evaluating its output.  For some, the replication step itself demanded this kind of reflection, because generating a useful practice problem required understanding the original one: \textit{``I had to actually research the problem to understand how to make it, so I needed to understand the category, type of problem, and all the steps involved to recreate one''} (FSC). Notably, several students framed AI's fallibility as productive: \textit{``With some AI models making minor errors, I am still forced to analyze the AI's process of solving the problem, which helps me to further learn the problem''} (VSU).  A smaller group emphasized \textit{efficiency and accessibility} (11.9\%), particularly the ability to get help immediately rather than \textit{``waiting until class time or office hours''} (PSU-BW) or \textit{``going through notes back and forth until it clicks''} (FSC).

Not all students reported a change. A minority (6.8\%) described \textit{limited or cautious AI use}: preferring textbooks, lecture slides, and instructors, distrusting AI output, or avoiding it altogether. As one student put it, \textit{``I personally despise AI and try to avoid it as much as possible and only use it as a last resort''} (PSU-BW). These responses are a reminder that the positive perceptions reported earlier were not universal.

\textit{Suggested improvements (C3).}---Students' suggestions closely mirrored the more mixed ratings of AI reliability (B5). The most frequently raised concern was \textit{accuracy and reliability} (34.2\%): students described calculation errors, inconsistent responses, and explanations that were \textit{``outright wrong and that made things further confusing''} (LU). These accounts are consistent with our recent evaluation of large language models on AP Physics free-response questions, which found high overall scores but persistent errors in calculation and in qualitative explanation~\cite{Paul2026APAI}. The second most common theme, \textit{better-calibrated explanations and interactive scaffolding} (29.7\%), is especially relevant to the design of P-A.I.R. Many students wanted guidance that would help them work through problems themselves. One asked AI to \textit{``lay the foundation to understanding a question"} (LU), while another wanted it to \textit{``ask us questions as it solves the problems to make sure we understand"} (PSU-BW).  Students also pointed to the need for stronger \textit{visual or multimodal explanations} (13.9\%), observing that current tools struggle both to interpret and to generate physics diagrams. This observation echoes the systematic difficulties with diagrams, graphs, vector directions, and three-dimensional reasoning identified in the same evaluation~\cite{Paul2026APAI}. A further 10.9\% emphasized the importance of \textit{alignment with instructor or course methods}, noting that \textit{``the method AI provides was not taught or related to what was taught in class''} (LU). Finally, 22.3\% explicitly stated that no improvement was needed.
 
\textit{Concerns beyond the codebook.}---Two concerns fell outside the codebook but warrant mention because they were raised spontaneously, without any survey item prompting them. First, four students from three of the four institutions raised the environmental cost of generative AI. One cited it as the reason for not using AI at all (\textit{``I don't use AI for environmental reasons''}, VSU).  Another, asked how the process could be improved, called for AI to be made \textit{``more ethical to use''}, expressing concerns about its energy consumption and its contribution to rising prices for computer memory, storage, and graphics processors (PSU-BW). Second, five students worried that reliance on AI could erode the cognitive work that learning physics requires. These concerns ranged from the blunt claim that AI \textit{``lowers brain function''} (LU) to the more measured observation that \textit{``relying too much on AI can reduce critical thinking and problem solving skills, which are essential in subjects like physics''} (LU). They also included warnings that answer-giving AI \textit{``will make people rely on AI too much and stagnate learning''} (PSU-BW), and that its efficiency could \textit{``increase procrastination for not all but some students like me''} (PSU-BW). Two of these students raised both environmental and cognitive concerns.
 
Although infrequent, these responses echo the risk of passive dependence that motivated P-A.I.R.\ in the first place~\cite{Paul2026PAIR}. They also align with students' own requests for scaffolding over answers: several respondents independently proposed the remedy the framework is built on. One wrote, \textit{``I should try and find my own mistake as to allow myself the opportunity to learn from it on my own''} (VSU). Taken together, the qualitative findings corroborate the quantitative results. Students primarily valued P-A.I.R.\ for helping them understand physics problems. Their concerns about accuracy, alignment with instruction, and over-reliance suggest priorities for improving its use in physics courses.

\section{Discussion and Conclusions}
\label{disc}
This study examined whether the positive student experiences reported in the single-institution pilot of P-A.I.R.~\cite{Paul2026PAIR} extend to a broader population of introductory physics students. Across four institutions, algebra- and calculus-based courses, and both semesters of the introductory sequence, the answer is largely yes. Students perceived clear conceptual and practice-related benefits (RQ1), these perceptions were consistent across instructional contexts (RQ2), and post-engagement confidence was associated with students' experience of the framework but not with how often they used AI (RQ3).

\textit{Perceived benefits (RQ1).}---The two components at the core of the framework received the strongest endorsement. Over 90\% of students reported improved conceptual understanding, and nearly all found the replicated practice problems helpful. The strong association between these items ($r_s = 0.58$) is consistent with the intended sequence of P-A.I.R., in which an AI-generated worked example reduces initial cognitive load~\cite{sweller1988cognitive} and the subsequent independent attempt reintroduces the retrieval and generation demands that consolidate learning~\cite{roediger2006test, slamecka1978generation}. The qualitative responses suggest how students experienced this sequence. The dominant theme was not that AI supplied answers, but that it helped students break problems into steps, decide which principles applied, and understand why a method worked. Smaller but meaningful groups of students described generating their own practice problems, attempting problems before consulting AI, and checking or critiquing AI output. In the terms of the ICAP framework~\cite{chi2014icap}, these are constructive and interactive forms of engagement, in contrast to the passive reading of solutions that unstructured AI use tends to encourage~\cite{kirschner2006minimal}. Several students described AI’s occasional errors as a prompt for closer analysis, illustrating how the reflection step can turn imperfect output into an opportunity for evaluation.

\textit{Consistency across institutions (RQ2).}---A central motivation for this study was to distinguish effects associated with the framework from those tied to a particular instructor, course, or student population. Mean ratings for conceptual understanding and replicated practice were broadly similar across institutions, post-engagement confidence varied only narrowly between institutions, and the dominant qualitative theme appeared at closely similar rates at every site. This consistency held despite differences in course level, student majors, survey administration, and local implementation details. Because the four-step protocol was the principal common element across these settings, the convergence suggests that the framework’s structure may contribute to similar student perceptions across varied classroom contexts.

\textit{Confidence and the role of AI experience (RQ3).}---Post-engagement confidence was most strongly associated with perceived conceptual clarity and the helpfulness of replicated practice, more weakly with prior confidence in using AI, and was essentially unrelated to AI usage frequency. Confidence was likewise not significantly related to self-reported GPA, suggesting that favorable confidence ratings were not limited to high-achieving students. Taken together, these patterns support the premise that motivated this work: \textit{how} students engage with generative AI matters more than \textit{how often} they use it. Frequent use alone did not translate into confidence in independent problem solving; productive engagement with explanation and practice did.

\textit{Implications for instruction.}---The concerns students raised also suggest specific ways to strengthen the framework. Concerns about accuracy were the most frequently raised issue and were consistent with the more mixed ratings of AI reliability. They also agree with independent evidence that current models, while generally strong on structured problems, remain prone to errors in calculation, qualitative explanation, and diagrammatic and three-dimensional reasoning~\cite{Paul2026APAI}. These concerns highlight the importance of keeping verification explicit within the protocol, so that checking AI output becomes an integral part of learning. A second group of suggestions called for explanations better matched to students' needs and for AI that supports reasoning without simply supplying answers. This is a pedagogically significant request: students independently asked for the kind of scaffolding that P-A.I.R.\ is designed to provide. One practical response is for instructors to provide prompt templates that direct the AI to guide students through a problem using questions and hints, and to follow the notation and solution methods used in the course. Such templates would address both students' preference for guidance and the mismatch some students noticed between AI-generated methods and those taught in class, while keeping students responsible for planning, monitoring, and evaluating their own learning, as emphasized in recent work on AI-supported self-regulated learning~\cite{LanZhou2025}.

Finally, a minority of students were skeptical of AI altogether. Some raised concerns that extended beyond the framework itself, including the environmental cost of generative AI and the possibility that reliance on it could weaken critical thinking and problem-solving skills. These responses mirror the distinction between pragmatic users and skeptical non-users reported in recent physics education research~\cite{Becker2026}, and they warrant serious consideration. The concern about reduced cognitive engagement is one that P-A.I.R.\ was designed to address: by asking students to attempt new problems independently before reviewing AI solutions, the framework aims to preserve their responsibility for the essential reasoning. The ethical and environmental concerns lie beyond what any instructional framework can resolve. Instructors can acknowledge these concerns by discussing the costs and limitations of generative AI openly in class and keeping the use of AI-supported frameworks such as P-A.I.R.\ optional, as it was in this study, so that students can choose an approach that aligns with their own values.

\textit{Limitations.}---These findings should be interpreted in light of several limitations. The study relies on self-reported perceptions and does not include direct measures of learning, and the absence of a pre-intervention measure or comparison group means that perceived gains cannot be attributed to P-A.I.R.\ with certainty. Participation in the survey was voluntary, and respondents’ experiences and perceptions may differ from those of non-respondents. Although the four-step protocol was common to all sites, implementation details, course structures, and survey administration (online versus paper) varied across institutions, and the frequency with which individual students applied the protocol was not consistently recorded across institutions. Finally, some qualitative themes were closely related, so assigning a response to one theme or another sometimes required judgment; independent coders initially disagreed on a number of such cases, and all were resolved through discussion before analysis.

\textit{Conclusions.}---Across four institutions and diverse student populations, students who engaged with the P-A.I.R.\ framework consistently reported improved conceptual understanding, found AI-generated replicated practice helpful, and expressed confidence in solving similar problems independently.  This confidence was linked to students’ perceived benefits from the framework, but not significantly to the frequency of their AI use, and was consistent across instructional contexts. Students' responses emphasized understanding over answer retrieval, while their concerns about accuracy, alignment with instruction, and over-reliance identify clear directions for improvement. As generative AI becomes more common in students’ study routines, instructors need approaches that support thoughtful use while accommodating students who choose not to use it. Frameworks that sequence AI assistance with independent problem solving offer a promising approach for encouraging students’ active participation in physics learning.

\acknowledgments{}
The authors extend their gratitude to their students for their participation in this study.

\bibliographystyle{unsrt}
\bibliography{main}

\end{document}